\def\cm{cm$^{-1}$}
\def\efa{Eu\-Fe$_2$\-As$_{2}$}

\def\BaFeCoxAs{Ba\-(Fe$_{1-x}$\-Co$_x$)$_2$\-As$_{2}$}

\def\bfca{Ba\-(Fe$_{0.92}$\-Co$_{0.08})_2$\-As$_{2}$}
\def\bfna{Ba\-(Fe$_{0.95}$\-Ni$_{0.05})_2$\-As$_{2}$}

\def\tsdw{$T_{\rm SDW}$}
\def\tc{$T_{c}$}
\def\EuFeAsP{Eu\-Fe$_2$\-(As$_{1-x}$\-P$_x$)$_{2}$}
\def\efap{Eu\-Fe$_2$\-(As$_{0.82}$\-P$_{0.18}$)$_{2}$}

\documentclass[prb,superscriptaddress,twocolumn,showpacs,amsmath,amssymb]{revtex4}
\usepackage[dvips]{graphicx}
\usepackage{graphicx}
\usepackage{color}

\begin{document}

\title{Optical investigations of the chemical pressurized Eu\-Fe$_2$\-(As$_{1-x}$\-P$_x$)$_{2}$:\\
an $s$-wave superconductor with strong interband interaction}

\author{D. Wu}
\affiliation{1.~Physikalisches Institut, Universit\"at Stuttgart, Pfaffenwaldring 57, 70550 Stuttgart, Germany}
\author{G. Chanda}
\affiliation{1.~Physikalisches Institut, Universit\"at Stuttgart, Pfaffenwaldring 57, 70550 Stuttgart, Germany}
\author{H. S. Jeevan}
\affiliation{I.~Physikalisches Institut,
Georg-August-Universit\"at G\"ottingen, 37077 G\"ottingen,
Germany}
\author{P. Gegenwart}
\affiliation{I.~Physikalisches Institut,
Georg-August-Universit\"at G\"ottingen, 37077 G\"ottingen,
Germany}
\author{M. Dressel}
\affiliation{1.~Physikalisches Institut, Universit\"at Stuttgart, Pfaffenwaldring 57, 70550 Stuttgart, Germany}

\date{\today}

\begin{abstract}
Superconducting Eu\-Fe$_2$\-(As$_{0.82}$\-P$_{0.18}$)$_{2}$ single crystals are investigated by infrared spectroscopy in a wide frequency range. Below $T_c=28$~K a superconducting gap forms at $2\Delta_{0} = 9.5~{\rm meV} = 3.8 k_B T_c$ causing the reflectivity to sharply rise to unity at low frequency. In the range of the gap the optical conductivity can be perfectly described by BCS theory with an $s$-wave gap and no nodes.
From our analysis of the temperature dependent conductivity and spectral weight at $T>T_c$, we deduce an increased $interband$ coupling between hole- and electron-sheets on the Fermi surface when $T$ approaches $T_c$.
\end{abstract}

\pacs{
74.25.Gz,    
74.70.Xa,    
74.25.Jb,    
74.20.Rp    
}
\maketitle

From the intense study of the new iron pnictides over the last years two main facts became clear. (1)~Optimum superconductivity is reached when long-range magnetic order and structural transition are completely suppressed by doping or pressure; magnetic excitations are responsible for superconductivity. (2)~Multi-band models are needed to describe the low-energy electronic excitations in pnictides and to account for the complexity and interplay of magnetic and superconducting orders; both intraband and interband interactions can play an important role.\cite{Mazin10,DavidRV10} To clarify the possible mechanisms, it is important to investigate the electrodynamics of charge carriers in compounds with an altered Fermi-surface environment.

The isovalent substituted $M$\-Fe$_2$\-(As$_{1-x}$\-P$_x$)$_{2}$ system ($M$=Ba, Eu, Sr, for instance) is an ideal candidate for this purpose because the chemical pressure induced superconductivity has a transition temperature as high as obtained by carrier doping.\cite{Renzhi09, Jiangshuai09} Band structure calculations show that phosphorous substitution significantly changes the shape of the hole sheets at the  $\Gamma$ point and finally one of the three hole sheets is absent in BaFe$_2$P$_2$.\cite{Shishido10, Kasahara10} Carrier doping (for example hole-doped Ba$_{1-x}$\-K$_x$\-Fe$_2$\-As$_2$), on the other hand, mainly affects the size of the sheets, though it influences the Fermi surface reconstruction as well.\cite{DingEPL08,Terashima09,Paglione10}
The isovalent substitution of As by P keeps the balance between the electron and hole numbers and avoids possible scattering effects by charged impurities within the Fe-As planes, which is still ambiguous in carrier doped systems.

Here we present an optical study on \efap\ in a wide temperature and energy range. When crossing \tc, clear evidence of a superconducting gap is observed at 74~\cm. The normalized conductivity resembles the BCS behavior with a complete gap 2$\Delta_{0}=3.8 k_B T_c$ without nodes. In the normal state, the spectral weight of the Drude component partially moves to the mid-infrared contributions upon cooling, indicating increasing interband coupling when the superconducting state is approached.

Single crystal samples of \efap were grown as described in Ref.~\onlinecite{Jeevan08}. Specific heat and magnetic susceptibility
measurements indicate a bulk superconducting transition at 28~K, followed by the magnetic ordering of Eu$^{2+}$ moments at 18~K.\cite{JeevanPdope} Shiny cleaved surfaces with size of 1$\times$1 mm$^2$ are available for optical experiments. The temperature dependent reflectivity spectra were measured in a wide frequency range from 20 to 37\,000~\cm\ using  two infrared Fourier-transform spectrometers (Bruker IFS 113v and IFS 66v/s) ($30-15\,000$~\cm) and a Woollam variable-angle spectroscopic ellipsometer extending up to the ultraviolet  ($6000 - 30\,000$~\cm, restricted to room temperature). Setups with two different 1.6~K bolometers (10-60~\cm\ and 20-100~\cm) have been employed to check and independently confirm the low-frequency reflectivity around the superconducting gap (below 100~\cm).

\begin{figure*}
 \centering
\includegraphics[width=1.9\columnwidth]{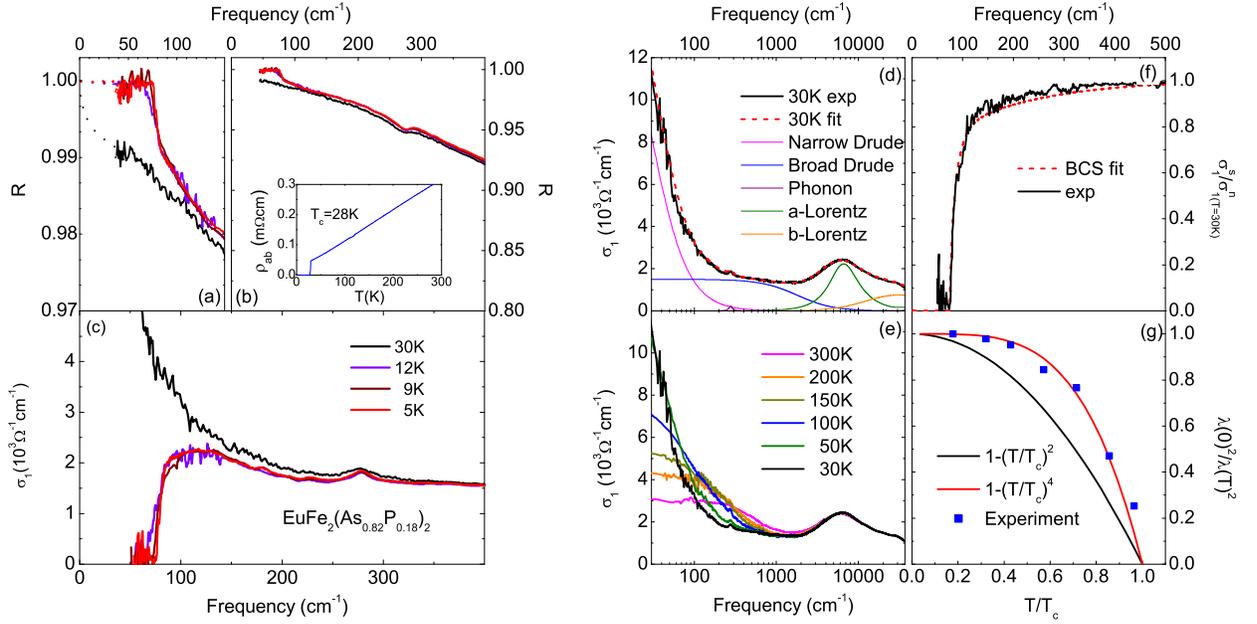}
 \caption{\label{fig:Fig1} (Color online)
Optical properties of \efap\ in the $ab$-plane measured in a broad frequency range at different temperatures. (a) and (b)~The reflectivity $R(\omega)$ increases with a change of curvature around 100~\cm\ for $T<28$~K and approach unity reflectance at about 70~\cm, which is a signature for a superconducting gap formation. Inset: in-plane $dc$-resistivity measured by four-contact method. The superconducting transition occurs at $T=28$~K, where the $\rho$ sharply decreases to zero. (c)~Below 28~K, the sudden drop of $\sigma_1(\omega)$ to zero indicates a nodeless $s$-wave superconducting gap. (d)~Drude-Lorentz fit for normal state conductivity ($T=30$~K). (e) In normal state, the optical conductivity develops as a metallic behavior upon cooling.(f) Ratio of the optical conductivity in the normal ($T=30$~K) and superconducting state $\sigma^{(s)}_1(\omega)/\sigma_1^{(n)}(\omega)$ of \efap\ at $T=5$~K.
The dashed red line corresponds to the fit by the BCS model using
a single gap of $2\Delta_0=74$~\cm.
(g)~Penetration depth as a function of temperature plotted in a renormalized fashion. The experimental data of \efap\ can be well described with $1-(T/T_c)^4$.}
\end{figure*}

In Fig.~\ref{fig:Fig1}(b) we show the in-plane reflectivity $R(\omega)$ of \efap\ at selected temperatures; panel (a) is a blow-up of the low-frequency part. The reflectivity has a metallic behavior both in frequency and temperature. Below \tc, an upturn with a change of curvature appears in R($\omega$) at around 100~\cm;  at the lowest temperature the reflectivity reaches unity at 74~\cm. This is strong evidence for the superconducting gap formation. The real part of optical conductivity $\sigma_1(\omega,T)$ calculated through Kramers-Kronig procedure is shown in panels (c) and (e) for different spectral ranges. Up to 4000~\cm, $\sigma_1(\omega,T>T_c$) can be described by a Drude contribution $\sigma_N$ plus a flat plateau which previously has been defined as a ``broad Drude'' $\sigma_B$.\cite{Wu09NP,Barisic10} For temperatures below \tc, $\sigma_1(\omega)$ drops rapidly to zero with decreasing frequency, indicating a nodeless $s$-wave superconducting gap. Two interband transitions are found at mid-infrared (labelled $\alpha$) and near-infrared ($\beta$) range, as demonstrated in Fig.~\ref{fig:Fig1}(d).

Multiple superconducting gaps have been reported in literature for the 122-family of pnictides,\cite{Dressel10} which is not surprising for   multi-band materials.\cite{DavidRV10} With this in mind, we also performed a BCS-based gap analysis for the optical properties of \efap\ to see whether we obtain a better description and gain information on the number of gaps and their magnitude. As mentioned above, two superimposed Drude components $\sigma_N$ and $\sigma_B$ are needed to describe the normal state conductivity of \efap. Correspondingly we tried to decompose the conductivity in the superconducting state $\sigma^{(s)}$ as $\sigma^{(s)}_{N}+\sigma^{(s)}_{B}$ assuming no interplay between them. In the simplest case, each of these independent components is described by the BCS theory\cite{Tinkham96,DresselGruner02} in the form given in Ref.~\onlinecite{Zimmermann91}. Surprisingly, the best description of $\sigma^{(s)}/\sigma^{(n)}$ is reached using the same gap parameter: $2\Delta_{N}=2\Delta_{B}=74$~\cm, as shown in Fig.~\ref{fig:Fig1}(f).
This implies a single $s$-wave gap 2$\Delta_{0}=3.8k_BT_c$ in \efap. Furthermore, it is not necessary to introduce nodes in the gap, as recently suggested\cite{Hashimoto10,Nakai10} from experiments on Ba\-Fe$_2$\-(As$_{1-x}$P$_x$)$_{2}$. Note, however, that we cannot rule out nodes that remain below our measuring range and accuracy. It is also possible, that  the Eu$^{2+}$ ions induced magnetic scattering that lifts the nodes. However, this suggestion needs further investigation.

From optical conductivity one can extract the superconducting penetration depth $\lambda(T)$ of \efap. According to Ferrell-Glover-Tinkham sum rule,\cite{Tinkham96,DresselGruner02}  the missing spectral weight $A= \int_{0+}^{\infty} \left[\sigma^{(n)}_1(\omega)- \sigma^{(s)}_1(\omega)\right]{\rm d}\omega$
is connected to the superfluid density  ${\rho_s}=8A/c^2$ and to the penetration depth as $A=c^2/{8{\lambda}^2}$. Alternatively, one can also estimate $\lambda(T)$ from the imaginary part of the complex conductivity in low-frequency limit:  $\sigma_2(\omega) = \frac{\pi N_s e^2}{ m \omega} = \frac{c^2}{4\pi \lambda^2 \omega}$. We consistently obtain $\lambda(T\rightarrow 0) \approx 300$~nm, which is very close to the value found for other 122 iron pnictides, such as \bfca\ and \bfna.\cite{Wu09NP} The temperature dependence of the penetration depth is plotted in Fig.~\ref{fig:Fig1}(g) in the form of ${\lambda(0)}^2/{\lambda(T)}^2$ {\it vs.} ${T}/{T_c}$. The data fall nicely onto the curve $1-\left({T}/{T_c}\right)^4$ as predicted by the two-fluid model and commonly used to describe a simple $s$-wave gap at moderate temperatures. These findings are consistent with our gap analysis and confirm the $s$-wave superconducting gap formation in \efap.

\begin{figure}
 \centering
\includegraphics[width=0.9\columnwidth]{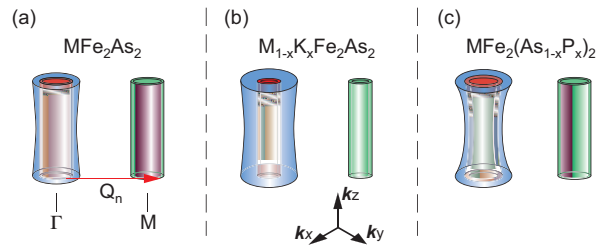}
 \caption{\label{fig:Fig3} (Color online)
Fermi sheets in first Brillouin zone for $M$Fe$_2$As$_2$  family ($M$ = Ba, Eu, Sr {\it et al.}). Three sheets at $\Gamma$ point are hole-like Fermi surfaces, and two at M point are electron-like Fermi surfaces. (a) Parent compound $M$Fe$_2$As$_2$; (b) hole doping leads to a change in the diameter of the Fermi surface in $M_{1-x}$K$_x$Fe$_2$As$_2$. The large gap $2\Delta_0=7.5 k_B T_c$ opens at the inner two hole-FSs (red sheets) and two electron-FSs (green sheets), while the small gap $2\Delta_0=3.7 k_B T_c$ opens at outer hole-FS (blue sheet). (c) P-substitution in $M$Fe$_2$(As$_{1-x}$P$_x$)$_2$ changes the warping of the Fermi surface and thus reduces the two dimensional character.}

\end{figure}

The most striking finding of our comprehensive investigations of 122 iron pnictides is that for the P-substituted system (chemical pressure) we observe a single superconducting gap, while for the carrier doped iron-pnictides multiple gaps are found.\cite{Li08SC,WuGAP10,Bernhard09,Dirk09,Dressel10}
To understand this difference, we have to consider the bandstructure, as depicted in Fig.~\ref{fig:Fig3}.
The undoped $M$Fe$_2$As$_2$ contains three sheets of the Fermi surface with hole character ($h$-FS) at the $\Gamma$ point and two electron sheets ($e$-FS) at the M point of the Brillion zone.\cite{Paglione10,Kasahara10,Zhang09}
While the inner two $h$-FSs and the two $e$-FSs are basically two-dimensional, the third $h$-FS is more three-dimensional, as it has a stronger dispersion along $k_z$.\cite{Fang10}
The antiferromagnetic (spin-density-wave) ordering happens along the nesting vector $Q_{n}$ indicated by arrow. With hole doping (K for instance), the $h$-FSs and $e$-FSs expand and shrink, respectively, along the $k_xk_y$ plane [Fig.~\ref{fig:Fig3}(b)], causing a considerable weakening of the nesting;  spin fluctuations increase. ARPES studies on optimally doped Ba$_{0.6}$K$_{0.4}$Fe$_2$As$_2$ show that the dispersion of the FSs remains similar to the undoped compounds; a large portion of the inner $h$-FSs can still be well connected to the $e$-FSs by the same $Q_{n}$, this enhances the scattering from the $e$-FSs to the $h$-FSs and yields a very large pairing strength ($2\Delta_{0}=7.5k_BT_c$) for these FSs.\cite{DingEPL08}

In P-substituted $M$Fe$_2$As$_2$, in the contrary, the isovalent substitution does not introduce carriers, hence no expansion or shrinking of FSs is expected. However, according to band structure calculations,\cite{Shishido10,Kasahara10,Analytis09} the P-substitution can enhance the dimensionality of $h$-FSs since it is very sensitive to the pnictogen position. As observed in recent ARPES investigations on P-substituted EuFe$_2$(As$_{1-x}$P$_x$)$_2$, the $k_z$ dispersion increases and varies along the $\Gamma$-Z line while it remains small and is slightly changed along the K-X line, namely the hole-FSs start to warp but eletron-FSs keep almost unchanged upon P-substitution.\cite{Thirupathaiah10} Such a warping of hole-FSs breaks the nesting condition and eventually enables superconductivity. At the same time, the warped FS rules out the possibility for a large geometric overlapping between the $h$- and $e$-FSs
that would cause a large pairing strength as discussed above for carrier-doped pnictides.
This peculiarity of the P-substituted systems can explain the absence of a $large$ superconducting gap in \efap. The gap pairing strength we observed in \efap\ is much closer to the smaller gap which opens at the outer $h$-FS in Ba$_{0.6}$K$_{0.4}$Fe$_2$As$_2$.\cite{DingEPL08} This could be related to the more three dimensionality of these Fermi sheets. Since the difference of $k_z$ dispersion between $h$- and $e$-FSs is more pronounced with the increasing dimensionality of $h$-FS, the scattering process near the $Q_{n}$ has to overcome the $k_z$ momentum, whereby reducing the pairing amplitude.

\begin{figure}
 \centering
\includegraphics[width=0.9\columnwidth]{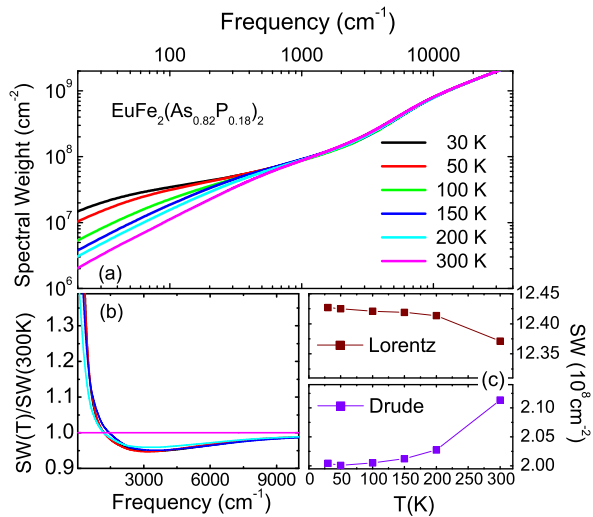}
 \caption{\label{fig:Fig4} (Color online)
(a) The spectral weight ${\rm SW}(\omega_c) = 8\int_0^{\omega_c}\sigma_1(\omega)\,{\rm d}\omega$ as a function of cut-off frequency $(\omega_c)$ for several temperatures. (b)~The normalized spectral weight ${\rm SW}(T)/{\rm SW(300~K)}$. A minimum can be found at $\omega_{\rm min}\approx 3000$~\cm. (c)~The spectral weight for the mid-infrared component $\alpha$-Lorentz and the Drude component plotted as a function of temperature for $T>T_c$.}
\end{figure}

More insight into the increased $interband$ coupling between hole- and electron-sheets on the Fermi surface (FS) is obtained from our spectral weight analysis, in particular when we consider the temperature dependence of the $\alpha$-interband contribution to the mid-infrared spectrum.
Optical studies
of electron-doped pnictides \cite{Barisic10,Degiorgi10} already reported some anomaly in this behavior, but the origin is not clear yet.
In the case of \EuFeAsP, the substitution of isovalent phosphorous always keeps the balance between hole and electron Fermi sheets; hence we can exclude scattering on charged impurities as suggested in \BaFeCoxAs.\cite{Degiorgi10} In order to do a more quantitative analysis, we present a spectral weight analysis that allows us to illustrate more precisely the development of Drude and Lorentz contribution with cooling. In Fig.~\ref{fig:Fig4}(a), the spectral weight ${\rm SW}(\omega_c) = 8\int_0^{\omega_c}\sigma_1(\omega)\,{\rm d}\omega$ is plotted as a function of cut-off frequency ${\omega_c}$. Note, up to 10\,000~\cm\ the ${\rm SW}(\omega_c,T)$ is still not conserved due to the temperature dependence of the $\alpha$-interband-contribution. More importantly, an upturn around $3000-5000$~\cm\ is observed, indicating that the spectral weight redistribution starts within this frequency range. To elucidate this point more clearly, we normalize ${\rm SW}(T<300~{\rm K})$ to ${\rm SW(300~K)}$ and plot it in Fig.~\ref{fig:Fig4}(b), and find that the spectral weight is reduced up to 5\%; an effect beyond the uncertainty of experiments. We also notice that, below the $\omega_{\rm min}$ (defined as a frequency where the relative spectral weight reaches its minimum), the normalized SW drops faster upon cooling as typical for a Drude behavior that narrows with decreasing temperature. Above $\omega_{\rm min}$ the ratio ${\rm SW}(T<300~{\rm K})$ to ${\rm SW(300~K)}$ increases as the interband transition starts to dominate the spectral weight.
We can easily extract the Drude and $\alpha$-Lorentz component by taking ${\rm SW_{Drude}}={\rm SW(\omega_{min})}$ and ${\rm SW_{\alpha-Lorentz}}={\rm SW_{cons}}-{\rm SW(\omega_{min})}$, as plotted in Fig.~\ref{fig:Fig4}(c); here ${\rm SW_{cons}}$ is the spectral weight approached for large frequencies, in practice $\omega\approx 10\,000$~\cm. With decreasing temperature, ${\rm SW_{Drude}}$ is continuously reduced and the lost ${\rm SW}$ piles up at the interband contribution. Normally, such a spectral weight shift from the Drude to interband transitions occurs when an energy gap forms in the density of states, as in the case of the parent compound \efa\ below \tsdw.\cite{Wu09}
A recent Hall-effect study on BaFe$_2$(As$_{1-x}$P$_x$)$_2$, however, rules out this possibility as it cannot explain the change of $R_H$ quantitatively.\cite{Kasahara10}
An alternative explanation of the reduced spectral weight is the enhancement of the effective carrier mass. In pnictides the electrons on the $e$-FSs scatter to the $h$-FSs via spin fluctuations. This is more effective due to band renormalization as both FS volumes shrink. Consistent with our picture, the shift of hole and electron bands towards the Fermi level is also observed in LaFeOP when taking the interband interaction into account.\cite{Ortenzi09} Therefore, we suggest that the spectral weight redistribution observed in our experiment is due to the interband interactions, and its enhancement with decreasing temperature indicates an increasing interband interactions when $T$ approaches \tc.

Our infrared study on isovalent P-substituted \efap\ ($T_c=28$~K) gives clear evidence for the formation of a single nodeless $s$-wave superconducting gap. From the comparison with the multi-gap scenario of carrier-doped iron pnictides, we found that the similarity in geometry and dimension between hole and electron Fermi-surface sheets has a strong influence not only on the nesting for the spin-density-wave state but also on the paring conditions for the superconducting state. This is considered as another indication that the pairing between the Fermi surface sheets very much relies on spin excitations along the nesting vector; in other words superconductivity in iron pnictide has magnetic origin. The continuous spectral-weight redistribution from the free-carrier (Drude) to the interband (Lorentz) contributions upon cooling indicates an increasing interband interaction as the superconducting transition is approached.

We thank D.N. Basov and S. Jiang for helpful discussions. The
contributions of J. Braun and S. Zapf to the
experiments are appreciated. D.W acknowledge her
fellowship by the Alexander von Humboldt Foundation.
Work at G\"ottingen supported by the DFG through the
SPP1458.


\begin{thebibliography}{99}
\bibitem{Mazin10}
I. I. Mazin, Nature \textbf{464}, 183 (2010).

\bibitem{DavidRV10}
D. C. Johnston, arXiv:1005.4392 (2010).

\bibitem{Renzhi09}
Z. Ren, Q. Tao, S. Jiang, C. M. Feng, C. Wang, J. H. Dai, G. H. Cao and Z-A. Xu, Phys. Rev. Lett. \textbf{102}, 137002 (2009)

\bibitem{Jiangshuai09}
S. Jiang, H. Xing, G. F. Xuan, C. Wang, Z. Ren, C. M. Feng, J. H. Dai, Z-A. Xu and G. H. Cao, J. Phys: Condens. Matter \textbf{21}, 382203 (2009).

\bibitem{Shishido10}
H. Shishido {\it et al.},
Phys. Rev. Lett. \textbf{104}, 057008 (2010).

\bibitem{Kasahara10}
S. Kasahara {\it et al.},
Phys. Rev. B \textbf{81}, 184519 (2010)

\bibitem{DingEPL08}
H. Ding {\it et al.},
Eur. Phys. Lett. \textbf{83}, 47001 (2008).

\bibitem{Terashima09}
K. Terashima {\it et al.},
PNAS \textbf{106}, 7330 (2009).

\bibitem{Paglione10}
J. Paglione and R. L. Greene, Nature Phys. \textbf{6}, 645 (2010)

\bibitem{Jeevan08}
H. S. Jeevan, Z. Hossain, D. Kasinathan, H. Rosner, C. Geibel, and P. Gegenwart, Phys. Rev. B {\bf 78}, 092406 (2008);
Phys. Rev. B {\bf 78}, 052502 (2008).

\bibitem{JeevanPdope}
H. S. Jeevan, D. Kasinathan, H. Rosner and P. Gegenwart, ({\it unpublished})

\bibitem{Wu09NP}
D. Wu {\it et al.},
Phys. Rev. B {\bf 81}, 100512(R) (2010).

\bibitem{Barisic10}
N. Barisic, D. Wu, M. Dressel, L. J. Li, G. H. Cao, and Z-A. Xu, Phys. Rev. B \textbf{82}, 054518 (2010).

\bibitem{Dressel10}
M. Dressel, D. Wu, N. Bari\v{s}i\'{c}, and B. Gorshunov, J. Phys. Chem. Solids (2010), and references therein.


\bibitem{DresselGruner02}M. Dressel and G. Gr\"uner, {\em Electrodynamics of Solids}
(Cambridge University Press, Cambridge, 2002).

\bibitem{Tinkham96}
M. Tinkham, \textit{Introduction to Superconductivity}, 2nd
edition (McGraw-Hill, New York, 1996).

\bibitem{Zimmermann91}
W. Zimmermann, E. H. Brandt, M. Bauer, E. Seider and L. Genzel, Physica C \textbf{183}, 99 (1991).

\bibitem{Hashimoto10}
K. Hashimoto {\it et al.},
Phys. Rev. B \textbf{81}, 220501 (2010).

\bibitem{Nakai10}
Y. Nakai, T. Iye, S. Kitagawa, K. Ishida, S. Kasahara, T. Shibauchi, Y. Matsuda and T. Terashima, Phys. Rev. B \textbf{81}, 020503 (2010).


\bibitem{Li08SC}
 G. Li, W. Z. Hu, J. Dong, Z. Li, P. Zheng, G. F. Chen, J. L. Luo, and N. L. Wang, Phys. Rev. Lett. \textbf{101}, 107004  (2008).

\bibitem{WuGAP10}
D. Wu, N.  Bari\v{s}i\'{c}, M. Dressel, G. H. Cao, Z. A. Xu, J. P. Carbotte and E. Schachinger, arXiv:1007.5215 (2010).

\bibitem{Bernhard09}
K. W. Kim, M. R{\"o}ssle, A. Dubroka, V. K. Malik, T. Wolf, and C. Bernhard, Phys. Rev. B {\bf 81}, 214508 (2010).

\bibitem{Dirk09}
E. van Heumen, Y. Huang, S. de Jong, A.B. Kuzmenko, M.S. Golden, and D. van der Marel, Europe. Phys. Lett. \textbf{90}, 37005 (2010).



\bibitem{Zhang09}
L. Zhang and D. J. Singh, Phys. Rev. B \textbf{79}, 174530 (2009).

\bibitem{Fang10}
G. T. Wang, Y. M. Qian, G. Xu, X. Dai and Z. Fang, Phys. Rev. Lett. \textbf{104}, 047002 (2010).

\bibitem{Analytis09}
J. G. Analytis, C. M. J. Andrew, A. I. Coldea, A. McCollam, J-H. Chu, R. D. McDonald, I. R. Fisher and A. Carrington, Phys. Rev. Lett. \textbf{103}, 076401 (2009).

\bibitem{Thirupathaiah10}
S. Thirupathaiah {\it et al.},
arXiv:1007.5205.

\bibitem{Degiorgi10}
A. Lucarelli, A. Dusza, F. Pfuner, P. Lerch, J. G. Analytis, J.-H. Chu, I. R. Fisher and L. Degiorgi, arXiv:1004.3022 (2010).

\bibitem{Wu09}
D. Wu {\it et al.},
Phys. Rev. B {\bf 79}, 155103  (2009).

\bibitem{Ortenzi09}
L. Ortenzi, E. Cappelluti, L. Benfatto and L. Pietronero, Phys. Rev. Lett. \textbf{103}, 046404 (2009).



\end{thebibliography}
\end{document}